\documentstyle[twocolumn,seceq,epsbox]{jpsj}

\def\runtitle{Perturbation Theory on the Transition Temperature 
and Electronic Properties of Organic Superconductor}
\def\runauthor{Takanobu {\sc Jujo}, Shigeru {\sc Koikegami} and Kosaku {\sc Yamada}}

\title
{
Perturbation Theory on the Transition Temperature 
and Electronic Properties of Organic Superconductor
}

\author
{ 
Takanobu {\sc Jujo}\footnote{E-mail: jujo@ton.scphys.kyoto-u.ac.jp}, 
Shigeru {\sc Koikegami} and Kosaku {\sc Yamada}
}

\inst
{
Department of Physics, Kyoto University, Kyoto 606-8502
}

\recdate{to be published in JPSJ (1999) No. 4}

\abst
{
We study the superconducting transition temperature and 
the electronic properties of the metallic phase of 
$\kappa$-type (BEDT-TTF)$_2$X which shows unconventional 
properties in experiments, on the basis of the third 
order perturbation theory for a simple effective 
Hubbard model of a nearly triangular lattice. 
Appropriate transition temperatures and $d_{x^2-y^2}$
symmetry of the gap function are obtained in good agreement with 
experimental results. We also calculate the transition temperature 
by the fluctuation-exchange approximation(FLEX) in order to compare the two 
approaches; FLEX gives higher transition temperatures 
rather than the perturbation approach. 
However, it is also found that the vertex corrections, 
which are ignored in FLEX, 
have a crucial effect on $T_{\rm c}$ for strongly frustrated systems. 
The density of states and the normal 
self-energy calculated in this perturbation scheme 
show the nature of the conventional Fermi liquid near the Mott-insulator. 
Thus, our perturbation approach is applicable to the conventional 
metallic phase of this compound, while it cannot explain 
the (pseudo-)spin gap phenomenon which signals the non-Fermi liquid.}

\kword
{
perturbation approach, electron correlation, 
spin fluctuation, vertex correction, Fermi liquid, 
d-wave superconductor, density of states, self-energy 
}

\begin{document}
\sloppy
\maketitle

\section{Introduction}
Today, there are many organic compounds which show the superconductivity 
at low temperatures. One of the most interesting 
superconductors among these compounds is  $\kappa$-type (BEDT-TTF)$_2$X 
(X=Cu(NCS)$_2$, Cu[N(CN)$_2$]Br, Cu[N(CN)$_2$]Cl and $\kappa$-(ET)$_2$X 
for an abbreviation). These compounds show 
superconductivity under the pressure appropriate to each compound. 
We have two ordered phases by varying pressure, one of which 
is antiferromagnetic insulator in the lower pressure region and the other is 
superconductivity in the higher pressure region.~\cite{rf:1} \par
By the nuclear magnetic resonance(NMR) experiments, 
it has been confirmed that, 
below the transition temperature($T_{\rm c}$), 
there are no coherence peaks, the Knight shifts decrease 
and the spin lattice relaxation rate behaves as $T^3$.~\cite{rf:2} 
These results imply that the symmetry of the Cooper pair 
is not the isotropic s-wave, but 
the anisotropic symmetry and spin singlet, indicating the d-wave. 
These properties are quite similar to the cuprate superconductors
(high-$T_{\rm c}$). The mechanism of this superconductivity 
cannot be explained on the basis of the conventional 
phonon-mediated superconductivity, which usually 
shows s-wave property and the isotope effect. \par 
Theoretically the unconventional superconductivity 
have been discussed keeping high-$T_{\rm c}$ cuprates in mind, 
roughly speaking from two approaches which are 
the theory based on the resonating-valence-bond(RVB) state~\cite{rf:3} 
($t$-$J$ model) and the spin-fluctuation model.~\cite{rf:4,rf:5,rf:6} 
In these approaches RVB theory assumes a strong on-site 
Coulomb interaction, and is expected to be effective 
to the doped Mott insulator like the high-$T_{\rm c}$ materials 
in the underdoped region. 
On the other hand, our system is metallic although it has the nature 
of the half-filling; this fact suggests that 
the on-site Coulomb interaction is not so strong. 
Therefore, we consider the weak-coupling approach 
is appropriate to this superconductor. 
When we discuss the spin-fluctuation mediated 
superconductivity, we have several methods 
of determining the attractive force. 
Moriya, {\it et al.} determined 
Im$\chi(\mbox{\boldmath $q$},\omega)$ (the imaginary part 
of the dynamical susceptibility) by the self-consistent 
renormalization(SCR),~\cite{rf:4} and Pines, {\it et al.} 
determined it by NMR experimental results.~\cite{rf:5} 
They solved the \'Eliashberg equation by using the 
phenomenologically determined susceptibility, and  
obtained the appropriate $T_{\rm c}$. 
Before them, simple calculations by the Random Phase Approximation
(RPA) were done by Shimahara {\it et al.}~\cite{rf:6} 
However, the $T_{\rm c}$ obtained by this method was rather lower 
than those of the above methods owing to the overestimated 
damping effect as pointed out by Hotta.~\cite{rf:7} 
(It should be noted that Shimahara also explored
the possibility of spin-fluctuation mediated superconductivity
in the organic superconductor (TMTSF)$_2$X.)  \par
These theories are expected to be effective 
to $\kappa$-(ET)$_2$X which shows strong antiferromagnetic 
spin fluctuation as observed by NMR experiments. 
Here, in order to understand these compounds 
on the basis of a more microscopic model, 
we take the following approach. \par
Because this superconductivity occurs only near 
the Mott-insulating phase under the controlled pressure, 
the electron correlation on the same site is expected 
to play an important role on the appearance of this 
superconductivity. One of methods to investigate 
electron correlation-induced superconductivity is 
the perturbation method, 
with respect to Coulomb repulsion $U$. 
In the studies of the transition temperature
for the high-$T_{\rm c}$ cuprates, 
this method was adopted by Hotta.~\cite{rf:7} 
He calculated $T_{\rm c}$ by the third order perturbation treatment(TOPT), 
and discussed the effect of the normal self-energy 
and the vertex correction on the calculated $T_{\rm c}$ 
and then found out that the vertex corrections suppress $T_{\rm c}$, 
compared with that obtained by a summation of the RPA-like diagrams.\par
Another method of investigating this kind of 
superconductivity from the weak coupling regime 
is the fluctuation exchange approximation(FLEX),~\cite{rf:8}
which is a kind of the self-consistent calculation 
based on the RPA-like diagrams. 
The use of this method is reasonable because 
the anisotropic superconductivities usually appear near 
the magnetic ordered phase by controlling 
the pressure or the doping of carriers. 
Therefore, the summation of the specific diagrams 
which are favorable to the spin fluctuation may be justified. 
However, in this work we concentrate on TOPT, without assuming 
the strong spin fluctuation like FLEX because it is 
not clear yet whether taking only the effect 
of particular diagrams is reliable for revealing the mechanism 
of superconductivity. \par
The perturbation approach is sensitive to the dispersion 
of the bare energy band by its nature, 
it implies that the lattice structures 
and the band filling play the essential roles in 
the calculation of $T_{\rm c}$ and the electronic properties. 
Although the TOPT calculation is already performed 
for the high-$T_{\rm c}$, it is important for the study of this compound 
to calculate $T_{\rm c}$ newly on the basis of the effective 
model of $\kappa$-(ET)$_2$X. \par 
In this paper we calculate $T_{\rm c}$ for the simple effective 
model by TOPT and investigate its $U$-dependence and the 
dependence on the effect of the frustration. 
Recently the calculations of $T_{\rm c}$ by FLEX were performed by 
Schmalian~\cite{rf:9} and Kino {\it et al.}, Kondo {\it et al.}~\cite{rf:10} 
on the two-band model and the single-band model respectively. 
Therefore we also calculate $T_{\rm c}$ by FLEX because these calculations 
were performed separately and there is no comparison with other 
methods yet; it is meaningful to compare the results by TOPT and FLEX. 
We also estimate the contribution of 
the vertex corrections, by comparing 
the results of the full TOPT calculation with that of only RPA-like 
terms included in anomalous self-energies. \par 
Another prominent property of this compound is the appearance 
of the (pseudo-)spin gap, which was revealed by the NMR 
experiments in the metallic phase.~\cite{rf:11}
Therefore, it is necessary to study 
the electronic properties in order to see what extent our model 
and approximation which are used to calculate $T_{\rm c}$ 
are pertinent to. We calculate the density of 
states and the normal self-energy, as the physical 
quantities characterizing the metallic phase of this compound, 
and then discuss the results and further problems. 
   
\section{Formulation}
\subsection{Hamiltonian}
First we define the effective Hamiltonian in a minimum form 
which describes $\kappa$-(ET)$_2$X for the calculation 
of $T_{\rm c}$. The Fermi surface(FS) of $\kappa$-(ET)$_2$X was studied 
by the Shubnikov-de Haas effect and found that it is well 
reproduced by the tight binding approximation based on 
the extended H\"uckel approximation.~\cite{rf:12} 
However it is known that a pair of ET molecules is considered 
as the basic structural unit(which is called a dimer) 
of the conduction sheet of $\kappa$-(ET)$_2$X 
because the intradimer hopping term is more than twice of the 
interdimer one.~\cite{rf:13,rf:14} 
Here, we briefly discuss the dimer model, which can be described 
as Fig.~\ref{fig:1}. 

\begin{figure}
\epsfile{file=fig1,height=10cm}
\caption{(a) The original lattice structure of $\kappa$-(ET)$_2$X when
BEDT-TTF molecules are written explicitly. The line represents 
each molecule. $t_d$ is the intradimer transfer integral. 
$t_1$, $t_2$ and $t_3$ constitute the interdimer transfer integral. 
(b) The lattice structure of $\kappa$-(ET)$_2$X when  
the pair of molecules is dimerized.
The circle represents a dimer, and the area which are 
surrounded by dashed lines, represent the unit cell 
in this case. 
$t$ and $t'$ are transfer integrals connecting the sites.}
\label{fig:1}
\end{figure}

The arrangement of BEDT-TTF molecules are shown in Fig. 1 (a). 
When a dimer is considered as a structural unit, 
the energy splitting between the bonding and antibonding orbitals 
is approximately given by $2t_d$.  
Therefore we can neglect the mixing between 
the bonding and antibonding orbitals, and the dimer model is verified. 
Focusing on the antibonding orbital, each dimer is connected 
to the nearest sites by two kinds of transfer integrals. 
These transfer integrals can be written as 
$t=(t_1+t_2)/2$ and $t'=t_3/2$. (The transfer integrals 
$t_1$, $t_2$ and $t_3$ correspond to those of Fig. 1 (a).) 
Thus, the lattice structure can be described as Fig. 1 (b) by 
using the above value of $t$ and $t'$. \par 
Strictly speaking, transfer integrals $ t $ in Fig. 1 
have two different values which cause the gap 
between the quasi-1D FS and quasi-2D FS 
and put two dimers in the unit cell. 
However, since the difference between them is quite small
(only 2 \% of its value),~\cite{rf:13} we consider its difference 
can be neglected and regard the minimum model as the lattice 
with one site in the unit cell. \par
For the reasons stated above we adopt the following Hubbard 
Hamiltonian 
\begin{eqnarray}
 {\cal H}=-t \sum_{<i,j>,\sigma}c^{\dag}_{i,\sigma}c_{j,\sigma} 
-t' \sum_{<i,k>,\sigma}c^{\dag}_{i,\sigma}c_{k,\sigma} \nonumber \\
+U \sum_{i}n_{i,\uparrow}n_{i,\downarrow}+{ \mu} \sum_{ \sigma}n_{i,\sigma} , 
\end{eqnarray}
where $\sigma$ is spin indices, $<i,j>$ indicates taking summation over 
nearest neighbor sites and $<i,k>$ over next nearest sites only 
in one direction as in Fig. 1. $\mu$ is the chemical potential which is 
determined so as to fix the electron number to 1 per site. 
The noninteracting energy dispersion is; 
\begin{equation}
  \epsilon_{ \mbox{\boldmath $k$}}=-2t(\mbox{cos}k_x+\mbox{cos}k_y)-2t'\mbox{cos}(k_x+k_y) .
\end{equation} 
According to the quantum chemistry calculations,~\cite{rf:15} 
we take $t'/t \simeq 0.7$ as the realistic value of $\kappa$-(ET)$_2$X. 
It should be noted that this lattice has strong frustration 
and can be called  a nearly triangular lattice. 
\subsection{Green's function and self-energy}
The non-interacting Green's function is the following; 
\begin{equation}
  G_0(\mbox{\boldmath $k$},\epsilon_n)=\frac{1}{{\rm i}\epsilon_n-(\epsilon_{ \mbox{\boldmath $k$}}-\mu)},
\end{equation} 
where $\epsilon_n=\pi T(2n+1)$ (n is integer) is the fermion-Matsubara frequency.
The diagrams of the normal self-energy are shown 
up to the third order of interaction in Fig.~\ref{fig:2} 
(the Hartree term is included in the shift of chemical potential). 

\begin{figure}
\epsfile{file=fig2,height=3.0cm}
\caption{The diagrams of the normal self-energy up to the third order. 
The solid and dashed lines correspond to the bare Green's function 
and the interaction, respectively.}
\label{fig:2}
\end{figure}

The analytic form of the normal self-energy is given by 
\begin{eqnarray}
  \Sigma_{\mbox{n}}(\mbox{\boldmath $k$},\epsilon_n)={\frac{T}{N}}\sum_{\mbox{\boldmath $k'$},n'}[U^2\chi_0(\mbox{\boldmath $k-k'$},\epsilon_n-\epsilon_{n'}) \nonumber\\
+U^3\chi_0^2(\mbox{\boldmath $k-k'$},\epsilon_n-\epsilon_{n'}) \nonumber\\
+U^3\phi_0^2(\mbox{\boldmath $k+k'$},\epsilon_n+\epsilon_{n'})]G_0(\mbox{\boldmath $k'$},\epsilon_{n'}).
\end{eqnarray}
In this equation $\chi_0$ and $\phi_0$ are represented as 
\begin{equation}
  \chi_0(\mbox{\boldmath $q$},\omega_m)=-{\frac{T}{N}}\sum_{\mbox{\boldmath $k$},n}G_0(\mbox{\boldmath $k$},\epsilon_n)G_0(\mbox{\boldmath $q+k$},\omega_m+\epsilon_{n}),
\end{equation}
\begin{equation} 
  \phi_0(\mbox{\boldmath $q$},\omega_m)=-{\frac{T}{N}}\sum_{\mbox{\boldmath $k$},n}G_0(\mbox{\boldmath $k$},\epsilon_n)G_0(\mbox{\boldmath $q-k$},\omega_m-\epsilon_{n}).
\end{equation}
The anomalous self-energy can be divided into two parts, 
one of which is included in FLEX, 
and the other is not included in FLEX, i.e. the vertex correction. 
These diagrams are shown in Fig.~\ref{fig:3} and Fig.~\ref{fig:4}, 
respectively. 

\begin{figure}
\epsfile{file=fig3,height=5.0cm}
\caption{The diagrams of the RPA parts(which are included in FLEX)
of the anomalous self-energy up to the third order.}
\label{fig:3}
\end{figure} 

\begin{figure}
\epsfile{file=fig4,height=5.0cm}
\caption{The diagrams of the vertex correction parts(which 
are not included in FLEX)
of the anomalous self-energy up to the third order.}
\label{fig:4}
\end{figure}

Analytically these diagrams are written as following equations. 
\begin{equation}
 \Sigma_{\mbox{a}}(\mbox{\boldmath $k$},\epsilon_n)=\Sigma_{\mbox{RPA}}(\mbox{\boldmath $k$},\epsilon_n)+\Sigma_{\mbox{vert}}(\mbox{\boldmath $k$},\epsilon_n),
\end{equation}

\vspace{5cm}

\begin{full}
\begin{equation}
\Sigma_{\mbox{RPA}}(\mbox{\boldmath $k$},\epsilon_n)=-{\frac{T}{N}}\sum_{k'}[U+U^2\chi_0(k+k')+2U^2\chi_0^2(k+k')]F(k'), 
\end{equation}
\begin{eqnarray}
\Sigma_{\mbox{vert}}(\mbox{\boldmath $k$},\epsilon_n)=-U^3{\frac{T^2}{N^2}}\sum_{k',k_1}G_0(k')(\chi_0(k+k')-\phi_0(k+k'))G_0(k+k'-k_1)F(k_1)  \nonumber \\
-U^3{\frac{T^2}{N^2}}\sum_{k',k_1}G_0(k')(\chi_0(-k+k')-\phi_0(-k+k'))G_0(-k+k'-k_1)F(k_1), 
\end{eqnarray}
\end{full}
where $k$=($\bf k$,$\epsilon_n$). 
\subsection{The particle number}
The particle number is $1.0$ per site in the real system. 
However, in our perturbation scheme, there is a discrepancy 
between the particle number based on the bare 
Green's function(which is denoted by $n_0$) 
and that based on the dressed Green's function(which is denoted by $n$) 
in our perturbation scheme. The bare 
susceptibility($\chi_0(\mbox{\boldmath $q$},\omega_m)$) plays 
an important role in the calculation of $T_{\rm c}$, that is, it 
determines the magnitude and the spatial and temporal variation 
of the interaction between the particles. To incorporate 
the nature of half-filling, we put $n_0=1$, and require the number 
conservation between $n_0$ and $n$. To satisfy this requirement 
we introduce the shift of chemical potential $\delta \mu$, which 
is included in the following dressed Green's function; 
\begin{equation}
  G(\mbox{\boldmath $k$},\epsilon_n)=\frac{1}{{\rm i}\epsilon_n-(\epsilon_{ \mbox{\boldmath $k$}}-\mu-\delta \mu+\Sigma_{\mbox{n}}(\mbox{\boldmath $k$},\epsilon_n))},
\end{equation}
and the particle number is given by 
\begin{equation}
 n = 2\frac{T}{N} \sum_{\mbox{\boldmath $k$},n}G(\mbox{\boldmath $k$},\epsilon_n).
\end{equation}
To satisfy $n = 1.0$ up to the third order of the interaction, 
expanding eq. (2.11) with regard to $\delta \mu - \Sigma_{\mbox{n}}(\mbox{\boldmath $k$},\epsilon_n)$
, and using $n_0=2 \sum_{\mbox{\boldmath $k$},n}G_0(\mbox{\boldmath $k$},\epsilon_n) = 1.0 $, 
then we get
\begin{equation}
 \delta \mu = - \frac{ \frac{T}{N}\sum_{\mbox{\boldmath $k$},n}G_0^2(\mbox{\boldmath $k$},\epsilon_n)\Sigma_{\mbox{n}}(\mbox{\boldmath $k$},\epsilon_n)}{\chi_0(\mbox{\boldmath $ 0 $},0)}.
\end{equation} 
\subsection{\'Eliashberg equation} 
For the calculation of $T_{\rm c}$, the Dyson-Gor'kov equation 
is linearized as 
\begin{equation} 
F(\mbox{\boldmath $k$},\epsilon_n)=|G(\mbox{\boldmath $k$},\epsilon_n)|^2\Sigma_{\mbox{a}}(\mbox{\boldmath $k$},\epsilon_n). 
\end{equation}
Then the \'Eliashberg equation is given by
\begin{full}
\begin{eqnarray} 
\Sigma_{\mbox{a}}(k)=-{\frac{T}{N}}\sum_{k'}[U+U^2\chi_0(k+k')+2U^2\chi_0^2(k+k')]|G(k')|^2\Sigma_{\mbox{a}}(k') \nonumber \\
-U^3{\frac{T^2}{N^2}}\sum_{k',k_1}G_0(k')(\chi_0(k+k')-\phi_0(k+k'))G_0(k+k'-k_1)|G(k_1)|^2\Sigma_{\mbox{a}}(k_1)  \nonumber \\
-U^3{\frac{T^2}{N^2}}\sum_{k',k_1}G_0(k')(\chi_0(-k+k')-\phi_0(-k+k'))G_0(-k+k'-k_1)|G(k_1)|^2\Sigma_{\mbox{a}}(k_1).
\end{eqnarray}
\end{full} 
When the eigen value of this equation is 1, the system is considered 
to be superconducting i.e. $ T =  T_{\rm c} $.
We don't assume the symmetry of the Cooper pair in calculating 
$T_{\rm c}$, which differs from Hotta,~\cite{rf:7} but calculating $T_{\rm c}$ and determining 
the symmetry are performed simultaneously. 

\section{Calculated Results}
\subsection{Details of the numerical calculation}
We take the transfer integral $ t $ as the unit of the 
energy and put $t=1$. From the quantum chemistry calculations,~\cite{rf:15}
$ t \simeq 70 \sim 80 $ meV, then $T_{\rm c}$ which is experimentally 
measured around 10K, is normalized to $T_{\rm c}=0.011$. 
The full bandwidth is from 8 to 9 when the value of $t'$ 
varies from 0 to 1. \par
To solve the \'Eliashberg equation we have to calculate 
the summations over the momentum and the frequency space. 
Since all these summations are in the convolution forms, 
we can carry out these by using the algorithm 
of the Fast Fourier Transformation.
For the frequency, irrespective of the temperature, 
we take $1024$ Matsubara frequencies, which makes 
the frequency cut-off comparable to the bandwidth at $T \simeq 0.0020$. 
On the other hand, for the momentum space, in order to 
avoid the finite size effect at low temperatures,~\cite{rf:16} 
we set points as fine as possible and divide the first 
Brillouin zone into $ 128 \times 128 $ meshes. 

\subsection{Transition temperature}
We calculate $T_{\rm c}$ by solving the \'Eliashberg equation $(2.14)$. 
The gap function shows the node at $k_x=k_y$ 
and $k_x=-k_y$ and changes the sign across the node 
in all approximations and for all parameters, namely  
\begin{equation} 
 \Sigma_{\mbox{a}}(\mbox{\boldmath $k$},\epsilon_n) \propto \mbox{cos}k_x-\mbox{cos}k_y .
\end{equation} 
The symmetry of Cooper pair is $d_{x^2-y^2}$.
The $U$-dependence of $T_{\rm c}$ is shown in Fig.~\ref{fig:5}. 

\begin{figure}
\epsfile{file=fig5,height=7.0cm}
\caption{The calculated $T_{\rm c}$. TOPT(RPA-only) in this figure 
means that only RPA-like diagrams of anomalous self-energies 
up to third order are included.(All normal self-energies are 
included up to the third order.) The diagonal transfer 
$t'=0.7$. The unit of energy is the transfer $t$.}
\label{fig:5}
\end{figure}

In this figure, we show also the results obtained by 
FLEX calculation and that for only the RPA-like 
diagrams(of anomalous self-energies) included in TOPT, 
in addition to TOPT calculation for comparison. 
This results indicate that 
for larger $U$ higher $T_{\rm c}$ are obtained commonly 
for all approximations of calculations. 
This is quite natural for our perturbation scheme 
and corresponds well to the experimental results which are measured 
by applying the pressure;~\cite{rf:17} the pressure increases 
the bandwidth $W$ and then $U/W$ is reduced. \par
We can see the differences between TOPT and FLEX in Fig. 5; 
$T_{\rm c}$ calculated by FLEX is higher than TOPT for moderate values 
of $U$. This is because RPA-like diagrams are included up to higher 
order in FLEX, and the spin-fluctuation is largely enhanced. 
This reasoning is supported by comparison between TOPT and TOPT
(RPA-only); the RPA-like diagrams of anomalous 
self-energies(${\rm eq}.\enspace (2.8)$) are responsible for attractive 
interaction, while the vertex corrections(${\rm eq}.\enspace (2.9)$, 
this term is not included in FLEX as noted 
above) are responsible for repulsive interaction and then 
have the effect of lowering $T_{\rm c}$  
(This was firstly pointed out by Hotta.~\cite{rf:7}). 
Therefore it can be said that the mechanism based on TOPT 
is the spin fluctuation induced one.
The another difference between the results of TOPT and FLEX is 
their $U$-dependences. The rate of increase of $T_{\rm c}$ as $U$ is varied 
in FLEX is slightly smaller than that of TOPT. This is probably 
due to the fact that the damping rate in FLEX is large because 
it is largely enhanced by the RPA diagrams, as well as the attractive 
interaction is enhanced. \par
There are many quantum chemistry calculations as mentioned above~\cite{rf:15} 
and the obtained values for the transfer integral $t'$ 
are roughly from 0.5 to 0.8, 
but it is difficult to know the real value of 
the transfer integral. Therefore it is meaningful to study 
the $t'$-dependence of $T_{\rm c}$. The calculated results are shown in 
Fig.~\ref{fig:6}. 

\begin{figure}
\epsfile{file=fig6,height=7.0cm}
\caption{The calculated $t'$-dependence of $T_{\rm c}$ in TOPT. The value of 
$U$ is fixed to $6.50$.}
\label{fig:6}
\end{figure}
    
The obtained highest value is $T_{\rm c}=0.06400$ for $t'=0.30$. For 
this value of $U=6.5$, and within our numerical accuracy, 
$T_{\rm c}$ is not obtained for $t' \geq 0.8$. 
To obtain $T_{\rm c}$ for this region of $t'$, it is needed 
to increase $U$. At $t'=0.00$, the Fermi surface is of perfect nesting and 
the bare susceptibility diverges. For this case 
we couldn't obtain the $T_{\rm c}$. \par
The calculated results of the static bare susceptibility 
(${\rm eq}.\enspace (2.5)$) are shown in Fig.~\ref{fig:7} for various values of $t'$. 

\begin{figure}
\epsfile{file=fig7,height=7.0cm}
\caption{The momentum dependence of the static bare susceptibility 
for various values of $t'$. These results are obtained for $T=0.0500$.}
\label{fig:7}
\end{figure}

For the typical value $t'=0.70$, the susceptibility 
has the incommensurate peak around $(\pi ,\pi)$ 
and asymmetric peak around $(-\pi,\pi)$. This result 
indicates the $d_{x^2-y^2}$ symmetry of the gap function. 
This momentum dependence of the bare susceptibility 
is similar to the result of Kondo, {\it et al.}~\cite{rf:10}
obtained by FLEX. Therefore it can be said that both of 
the approaches which are TOPT and FLEX use 
the nesting property to acquire the spin fluctuation and 
the attractive interaction. 
Here, the $t'$-dependence has the following feature. 
When $t'$ is decreased, the peaks at this momenta become sharper 
and then at $t'=0.10$ the incommensurability and 
the asymmetry of the peaks disappear. These facts explain 
the results shown in Fig. 6, which generally shows 
higher $T_{\rm c}$ for smaller $t'$. Physically this 
corresponds to the fact that for smaller $t'$, the lattice is 
closer to the square lattice and the antiferromagnetic fluctuation 
is larger due to reduced frustration. 
$T_{\rm c}$ at $t'=0.1$ is lower than that of $t'=0.3$ in spite of 
its sharper peak for $t'=0.1$. This is 
because as $t'$ decrease, the nesting property is enhanced
and then the spectral density in the vicinity of 
the Fermi level moves to the two incoherent 
parts(i.e. which become the higher and the lower Hubbard bands
in the Mott insulator). 
This transformation of the spectrum is shown in Fig.~\ref{fig:8}

\begin{figure}
\epsfile{file=fig8,height=7.0cm}
\caption{The density of states as $t'$ is varied, at $U=6.50$ 
and $T=0.06500$. The inset shows this figure focused 
near the Fermi level.}
\label{fig:8}
\end{figure}

In order to see how the vertex corrections influence 
$T_{\rm c}$ when $t'$ is varied, we also calculate $T_{\rm c}$ by 
including only RPA-like diagrams(Fig. 3) in anomalous 
self-energies, in other words, neglecting the vertex 
corrections. The results are shown in Fig.~\ref{fig:9}. 

\begin{figure}
\epsfile{file=fig9,height=7.0cm}
\caption{$t'$-dependences of $T_{\rm c}$, calculated by including
only RPA-like terms in anomalous self-energies(All normal self-energies 
up to the third order are included.), and by TOPT. 
The value of $U$ is fixed to 6.50.}
\label{fig:9}
\end{figure}

\begin{figure}
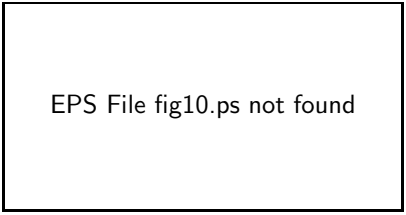

\epsfile{file=fig10,height=7.0cm}
\caption{$t'$-dependences of $|T_{\rm c}-T_{\rm c}^{\rm max}|/T_{\rm c}^{\rm max}$, 
these values are calculated based on Fig. 9. 
Here, $T_{\rm c}^{\rm max}$ is the maximum value of $T_{\rm c}$ 
by each method(i.e. TOPT or TOPT(RPA-only)), 
when $t'$ is varied. 
The value of $U$ is fixed to 6.50.}
\label{fig:10}
\end{figure}

For a comparison with TOPT in the order of 
magnitude, we take the linear-log plot as is shown 
in the figure. For the case that only RPA-like anomalous 
self-energies are included, we can also see that 
for smaller $t'$, higher $T_{\rm c}$ is obtained 
for the same reason as the calculation by TOPT(Fig. 6). 
However, as $t'$ is increased, 
the rate of decrease of $T_{\rm c}$ by TOPT
(which includes the vertex corrections) is larger 
than TOPT(RPA-only). This property of $t'$-dependence 
is clearly shown in Fig.~\ref{fig:10}. 

From Fig. 9 and Fig. 10, we can clearly see that for larger 
$t'$, $T_{\rm c}$ is largely reduced by taking the vertex corrections. 
At $t' \simeq 0.3, \enspace 0.4$, the value of $T_{\rm c}$ by TOPT 
is half or one third of that by TOPT(RPA-only). This ratio 
is comparable to the high-$T_{\rm c}$ calculation by Hotta.~\cite{rf:7} 
However, at $t'=0.70$ corresponding to the real system, the value of $T_{\rm c}$ by TOPT
is reduced by one order of magnitude compared with TOPT(RPA-only). 
This fact suggests that for large $t'$, i.e. when the frustration 
is large like our system, neglecting the vertex corrections 
is not a good approximation. 
Although the reason is not so simple from the complicated 
form of ${\rm eq}. (2.9)$, this property is probably due to the fact 
that the peaks of bare susceptibility are not so sharp 
for larger $t'$ as is shown in Fig. 7. Because ${\mbox R}{\mbox e}\phi _0(\mbox{\boldmath $q$},0)$ 
doesn't have large $t'$-dependence as is shown in Fig.~\ref{fig:11}, 
$\chi _0(\mbox{\boldmath $q$},0) - {\mbox R}{\mbox e}\phi _0(\mbox{\boldmath $q$},0)$ 
has more remarkable peak around $(0,0)$ for large $t'$ as is shown in Fig.~\ref{fig:12}. 
 
\begin{figure}
\epsfile{file=fig11,height=7.0cm}
\caption{The momentum dependence of ${\mbox R}{\mbox e}\phi _0(\mbox{\boldmath $q$},0)$
from ${\rm eq}. \enspace (2.6)$
for various values of $t'$. These results are obtained for $T=0.0500$.}
\label{fig:11}
\end{figure} 

\begin{figure}
\epsfile{file=fig12,height=7.0cm}
\caption{The momentum dependence of 
$\chi _0(\mbox{\boldmath $q$},0) - {\mbox R}{\mbox e}\phi _0(\mbox{\boldmath $q$},0)$
for various values of $t'$. These results are obtained for $T=0.0500$.}
\label{fig:12}
\end{figure}

From Fig. 12, it is seen that for small $t'$, 
the peak around $(\pi , \pi)$ makes 
the whole structure featureless. However, for large $t'$, 
the strong frustration makes the peak around $(0,0)$ remarkable.
Therefore the vertex corrections affect the gap function 
more repulsively for the d-wave symmetry at large $t'$ 
from ${\rm eq}. \enspace (2.9)$.
In other words, they suppress the antiferromagnetic spin-fluctuation. 
Although within the third order perturbation, it can be said 
that the vertex corrections have a crucial effect 
on the calculation of $T_{\rm c}$ for strongly frustrated systems. 
 
\subsection{Density of states and self-energy}
\begin{figure}
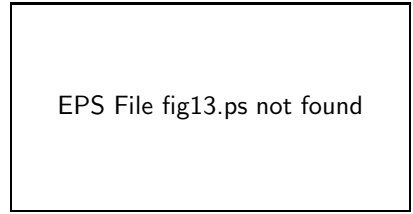

\epsfile{file=fig13,height=7.0cm}
\caption{DOS at $T=0.01500$, $t'=0.70$ and various values of $U$ 
as shown in the figure. The arrows indicate the incoherent peaks 
for each value of $U$.}
\label{fig:13}
\end{figure}

The density of states(DOS) is given by 
\begin{equation}
\rho (\omega ) = - \frac{1}{\pi }\sum_{\mib k}{\mbox I}{\mbox m}G(\mbox{\boldmath $k$},\omega). 
\end{equation}
We calculate ${\mbox I}{\mbox m}G(\mbox{\boldmath $k$},\omega)$ from 
${\rm eq}. \enspace (2.10)$ by using the Pad\'e approximation.~\cite{rf:18} 
The $U$-dependence of DOS is shown in Fig.~\ref{fig:13}. 

\begin{figure}
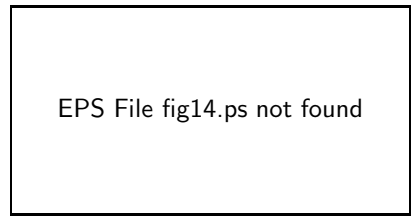

\epsfile{file=fig14,height=7.0cm}
\caption{DOS calculated by FLEX at $T=0.02000$, $t'=0.70$ 
and various values of $U$ as shown in the figure.}
\label{fig:14}
\end{figure}

From this figure we can see that the incoherent peaks, which are 
corresponding to the upper and the lower Hubbard 
peaks, grow as $U$ increases.(The upper peak can be seen for small $U$, 
while the lower peak can be seen only for large $U$. 
This tendency reflects the original asymmetric structure of DOS.)
This is the expected behavior for 
our perturbation scheme because for larger $U$ we approach the 
Mott-insulator phase, and coincides with the investigation on the 
infinite dimensions.~\cite{rf:19} 
This behavior cannot be obtained within FLEX as is shown in Fig.~\ref{fig:14}, 
although $U$ is very large and the system is expected
to be close to the Mott insulator.

The origin is probably ascribed to the self-consistent requirement in FLEX. 
In addition to this 
we see the asymmetric structure of DOS as expected 
for the nearly triangular lattice in Fig. 13
(i.e. because of the particle-hole asymmetry.). 
There is the coherence quasi-particle peaks near the Fermi levels 
for all values of $U$, and we cannot obtain the 
pseudogap behavior within our model and approximation. \par
The normal self-energy is obtained from ${\rm eq}. \enspace (2.4)$ 
by using Pad\'e approximation.~\cite{rf:18} 
The real part(${\mbox R}{\mbox e}\Sigma _n({\mib k},\omega )$) 
and the imaginary part(${\mbox I}{\mbox m}\Sigma _n({\mib k},\omega )$) 
of the self-energy at the Fermi momentum 
are shown in Fig.~\ref{fig:15} and Fig.~\ref{fig:16} respectively. 

\begin{figure}
\epsfile{file=fig15,height=7.0cm}
\caption{The real part of the normal self-energy 
at the Fermi momentum
,at $T=0.01500$, $t'=0.70$ and various values of $U$ 
as shown in the figure.}
\label{fig:15}
\end{figure}

\begin{figure}
\epsfile{file=fig16,height=7.0cm}
\caption{The imaginary part of the normal self-energy 
at the Fermi momentum
,at $T=0.01500$, $t'=0.70$ and various values of $U$
as shown in the figure}
\label{fig:16}
\end{figure}

The $\omega $-dependence of both parts near $\omega =0$ are 
respectively given by 
${\mbox R}{\mbox e}\Sigma _n({\mib k},\omega ) \propto - \omega $ and 
${\mbox I}{\mbox m}\Sigma _n({\mib k},\omega ) \propto - \omega ^2 $. 
(The small dip of the imaginary part at $U=7.0$ 
in Fig. 16, might be ascribed to the lack of accuracy of 
the Pad\'e approximation.)  
This behavior is the expected one for the usual Fermi liquid, 
and is opposite to the result of Maly, {\it et al.}~\cite{rf:20} which shows 
in their research in high-$T_{\rm c}$ 
$\left. \frac{\partial {\mbox R}{\mbox e} \Sigma _n({\mib k},\omega)}{\partial \omega}\right|_{\omega =0} > 0$ and $\left. \frac{\partial ^2{\mbox I}{\mbox m} \Sigma _n({\mib k},\omega)}{{\partial \omega}^2}\right|_{\omega =0} > 0$ obtained by using $t$-matrix approximation, resulting 
in the appearance of the pseudogap behavior. 
Other points which should be noted in Fig. 15 and Fig. 16 are 
that, as $U$ increases, the slope of ${\mbox Re}\Sigma _n({\mib k}_F,\omega )$ 
at $\omega =0$ becomes steeper and the coefficient of $\omega ^2$ 
for the imaginary part becomes bigger. 
This indicates that the mass and the damping rate of the quasi-particle 
become larger(physically this corresponds to the increase of 
the resistivity) as $U$ increases because the mass 
enhancement factor is given by $1-\left. \frac{\partial {\mbox Re}\Sigma _n({\mib k},\omega)}{\partial \omega}\right|_{\omega =0}$ and the damping rate is given by 
$-{\mbox Im}\Sigma _n({\mib k},\omega )$. These results are 
the typical Fermi liquid ones, and cannot show the anomalous 
behavior like the pseudogap. 
 
\section{Summary and Discussion}
In this paper we have calculated $T_{\rm c}$ 
for the simple effective model of 
$\kappa$-(ET)$_2$X by solving the \'Eliashberg equation 
on the basis of TOPT, and have investigated the possible mechanism 
by the spin fluctuation on this superconductor. 
We have adopted the Hubbard model with a nearly triangler lattice in 
which one of these transfer integrals, 
$t'=0.7$; we have taken $U$ as 
a parameter corresponding to the experiment in which 
the pressure is applied, and fixed the system to half-filling 
in agreement with the real compound. The obtained results roughly 
agree with the experiment in the high pressure region 
in its $U$-dependence and values. \par 
The results obtained here show that the superconductivity 
in this organic conductor possibly arises from the same mechanism, 
by which we mean the antiferromagnetic spin fluctuation induced 
superconductivity, as that of the cuprate system, 
since both of them are connected by changing a single parameter 
$t'$. \par
We have also compared the results obtained by TOPT and FLEX. 
For the moderate value $U$, $T_{\rm c}$ of FLEX is higher than 
that of TOPT. This is because FLEX involves the higher 
order spin fluctuation in spite of its large damping rate. 
However, from our perturbative aspect, because of the offset 
between large attractive interaction and large damping which 
are both given by RPA-like diagrams, the appropriate values of 
$T_{\rm c}$ are obtained by FLEX. Therefore it should be noted 
that whether so large enhancement is physical one or not
is a future problem. Concerning this point, we have shown 
in Fig. 5 and Fig. 9 that the vertex corrections reduce $T_{\rm c}$ by 
one order of magnitude. Up to the third order terms, it is 
found that the vertex corrections reduce $T_{\rm c}$ 
of strongly frustrated systems like our system, 
more seriously than that of not so frustrated systems 
like high-$T_{\rm c}$, although it cannot be said generally beyond 
the third order from our scheme.  
This fact suggests that the calculations 
of $T_{\rm c}$ which include only the RPA-like terms are questionable 
and should be carefully performed with the vertex 
corrections \par 
To investigate to what extent this scheme is effective, 
we study DOS and the normal self-energy. 
The obtained behaviors of both quantities are 
the expected ones for the Fermi liquid near the 
Mott-insulating phase. Especially the upper and the lower 
Hubbard bands in DOS are obtained in our perturbation scheme, 
while they are not obtained by FLEX. 
From this behavior of DOS, 
it is doubtful that the physical quantities obtained by FLEX 
have the nature of being near the Mott-insulating phase. 
It can be said that one of the elements which improve 
FLEX is possibly to take this property into account. \par
From the results of DOS and the self-energy, 
this perturbation approach is 
valid in the conventional metallic phase in the phase 
diagram of McKenzie.~\cite{rf:21} 
However the (pseudo-)spin gap which is one of the most 
interesting properties in the strongly correlated electron 
system, is not reproduced in our scheme. \par
From the analytic property in the power series 
expansion of $U$ for the Fermi liquid, 
we cannot derive this unconventional 
property by using the present method like TOPT or FLEX. 
There must be the crossover which shows the breakdown 
of the Fermi liquid at a certain value of $U$ as is 
already found in experiments. 
It has not been known yet that this breakdown is caused by 
the superconducting fluctuation~\cite{rf:20,rf:22} 
or the antiferromagnetic long(or short)-range order~\cite{rf:5,rf:23} or 
other mechanism. It is important to decide the critical term which 
characterizes the crossover and its effective region 
of the temperature and the magnitude of correlation. 
This is the problem to be solved 
to understand this compound and the nature of the 
non-Fermi liquid adjacent to the Fermi liquid. 

\section*{Acknowledgment}
Numerical computation in this work was carried out at the 
Yukawa Institute Computer Facility.

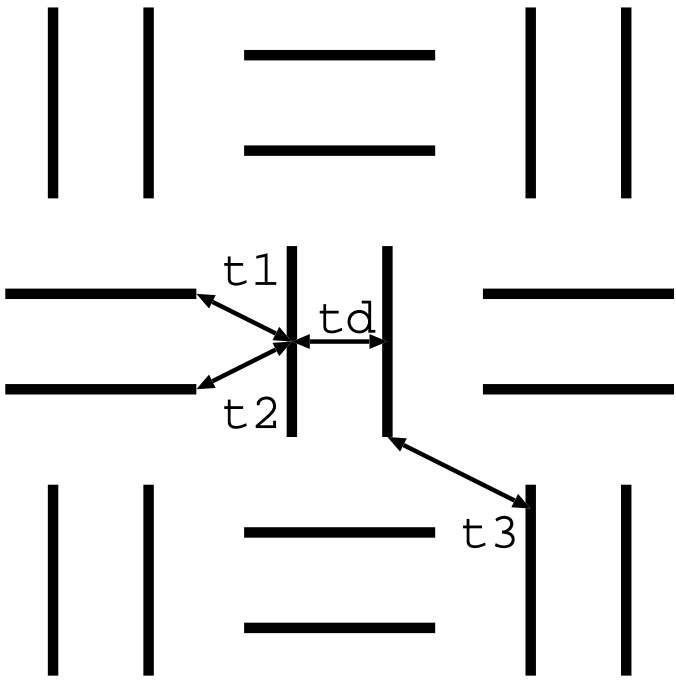
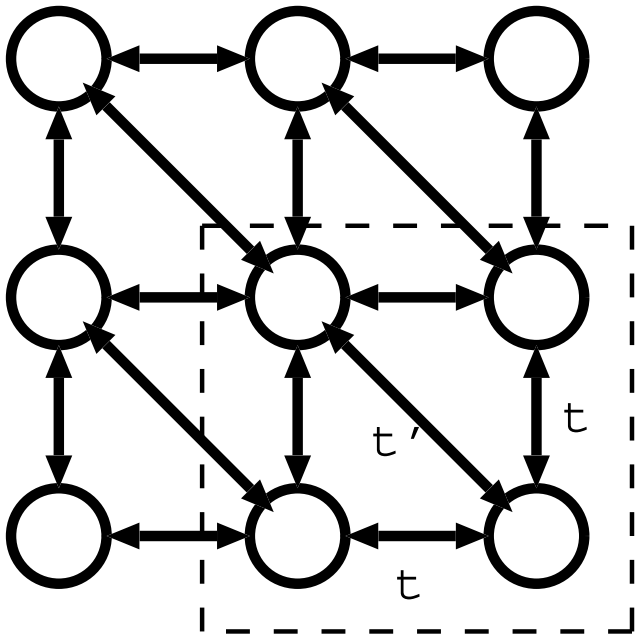
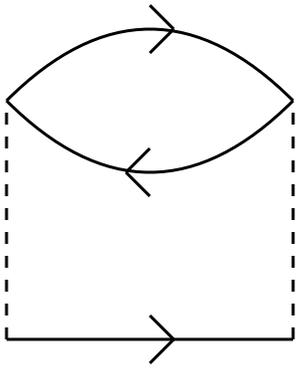
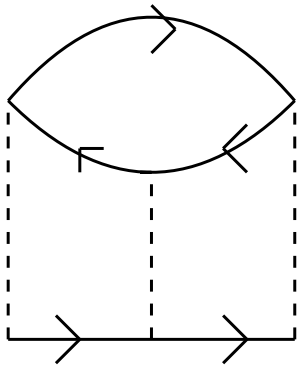
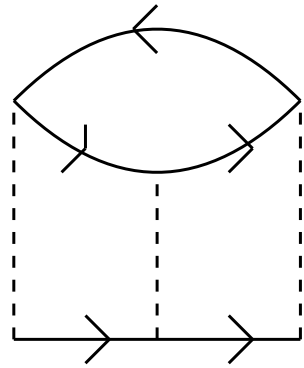
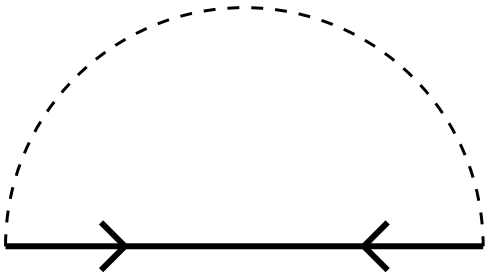
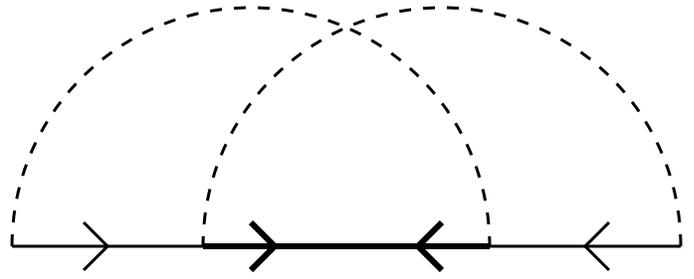
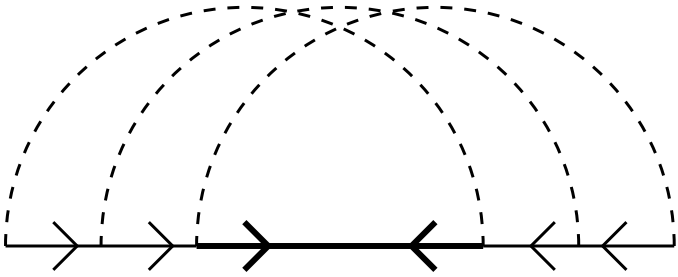
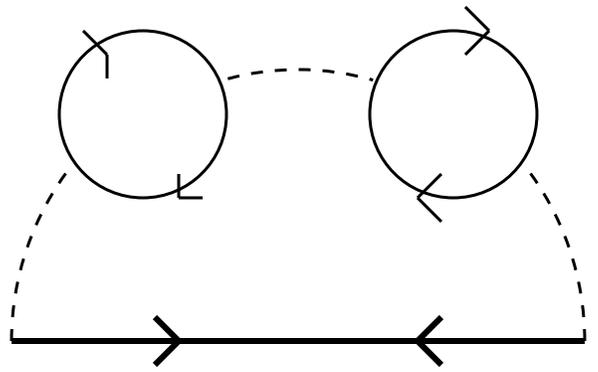
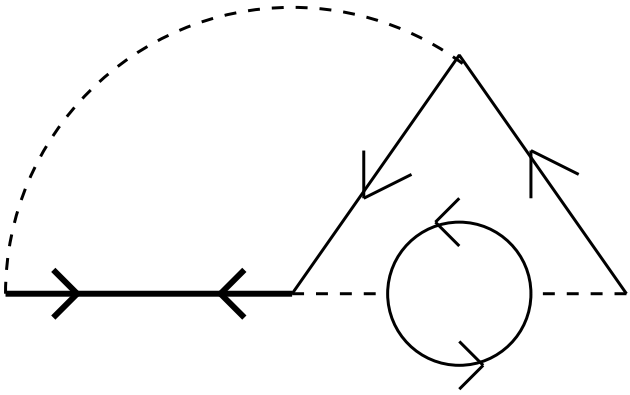
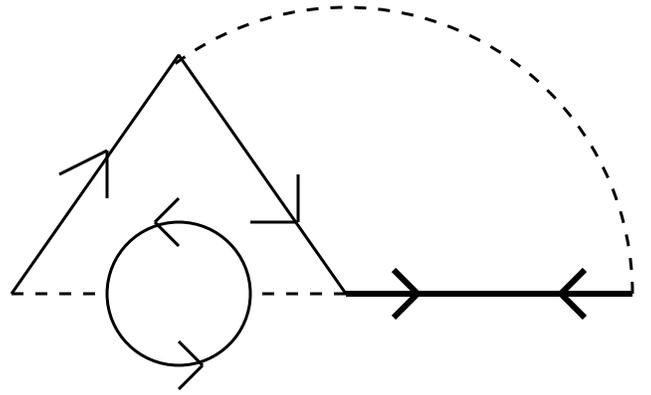
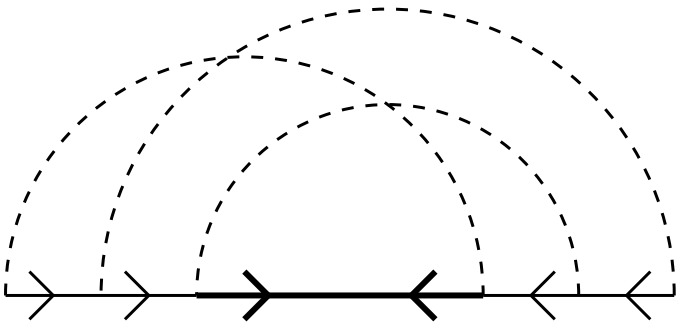
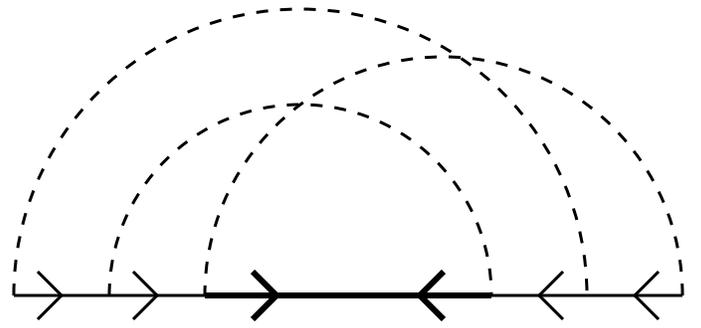
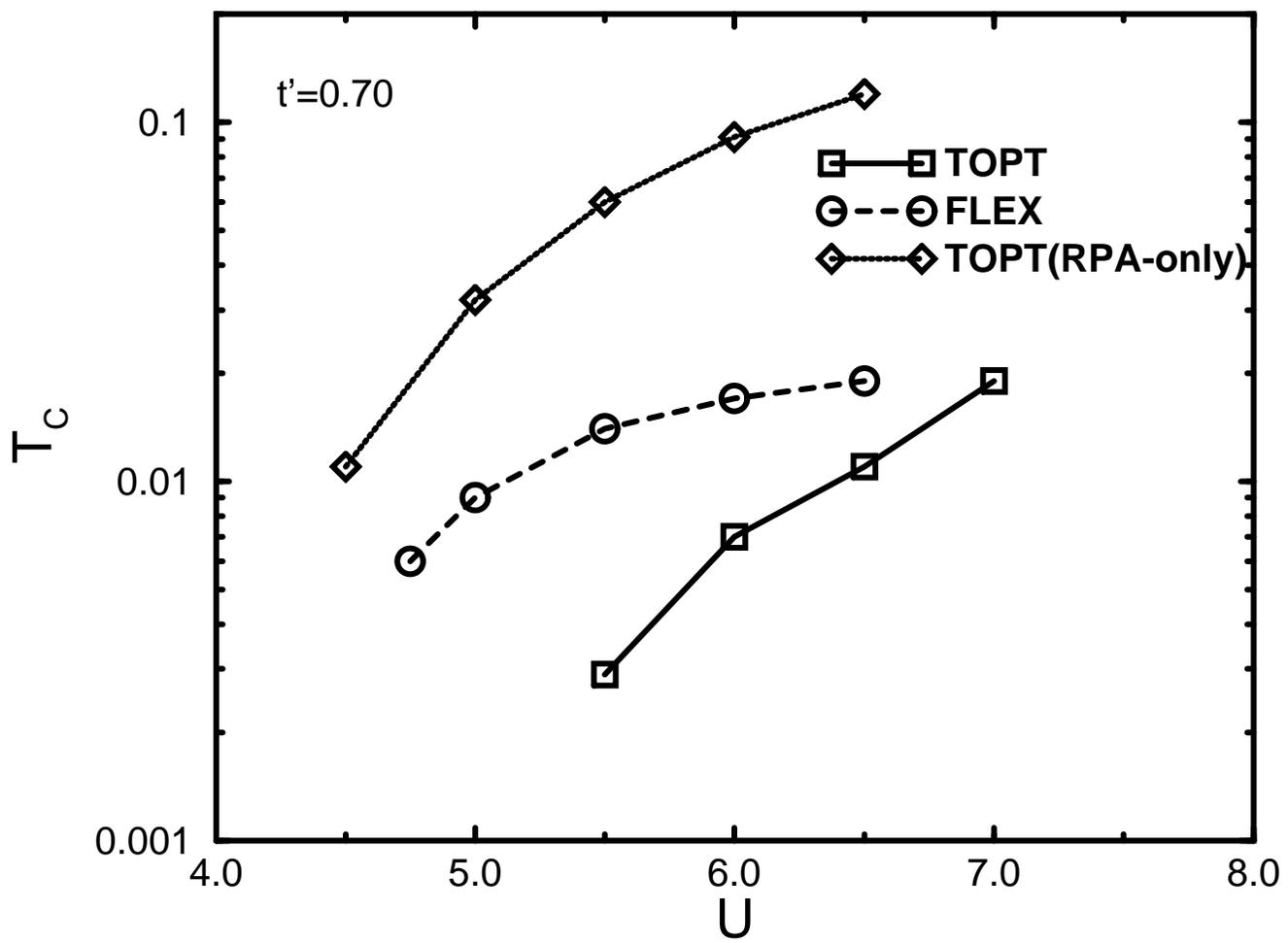
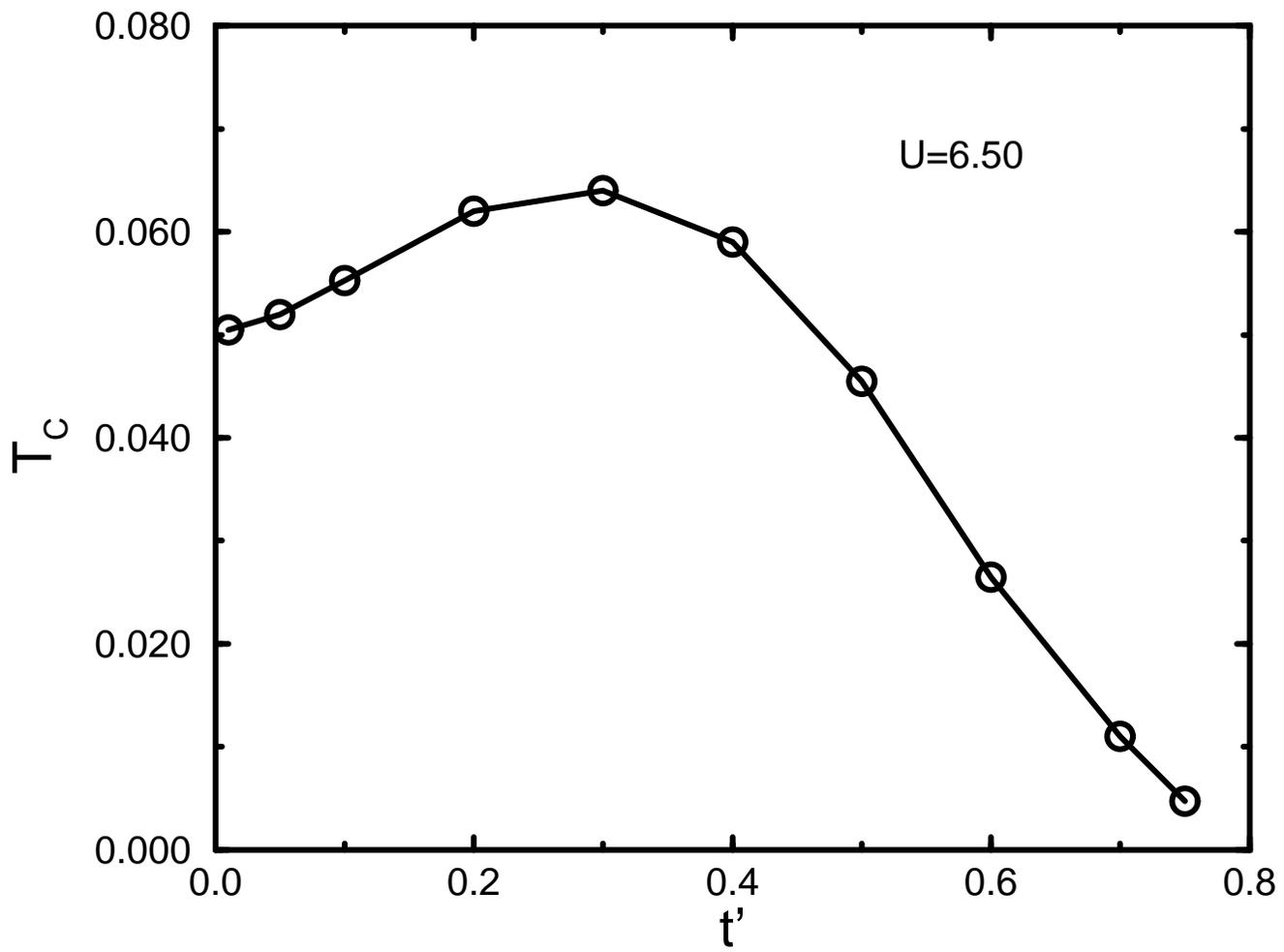
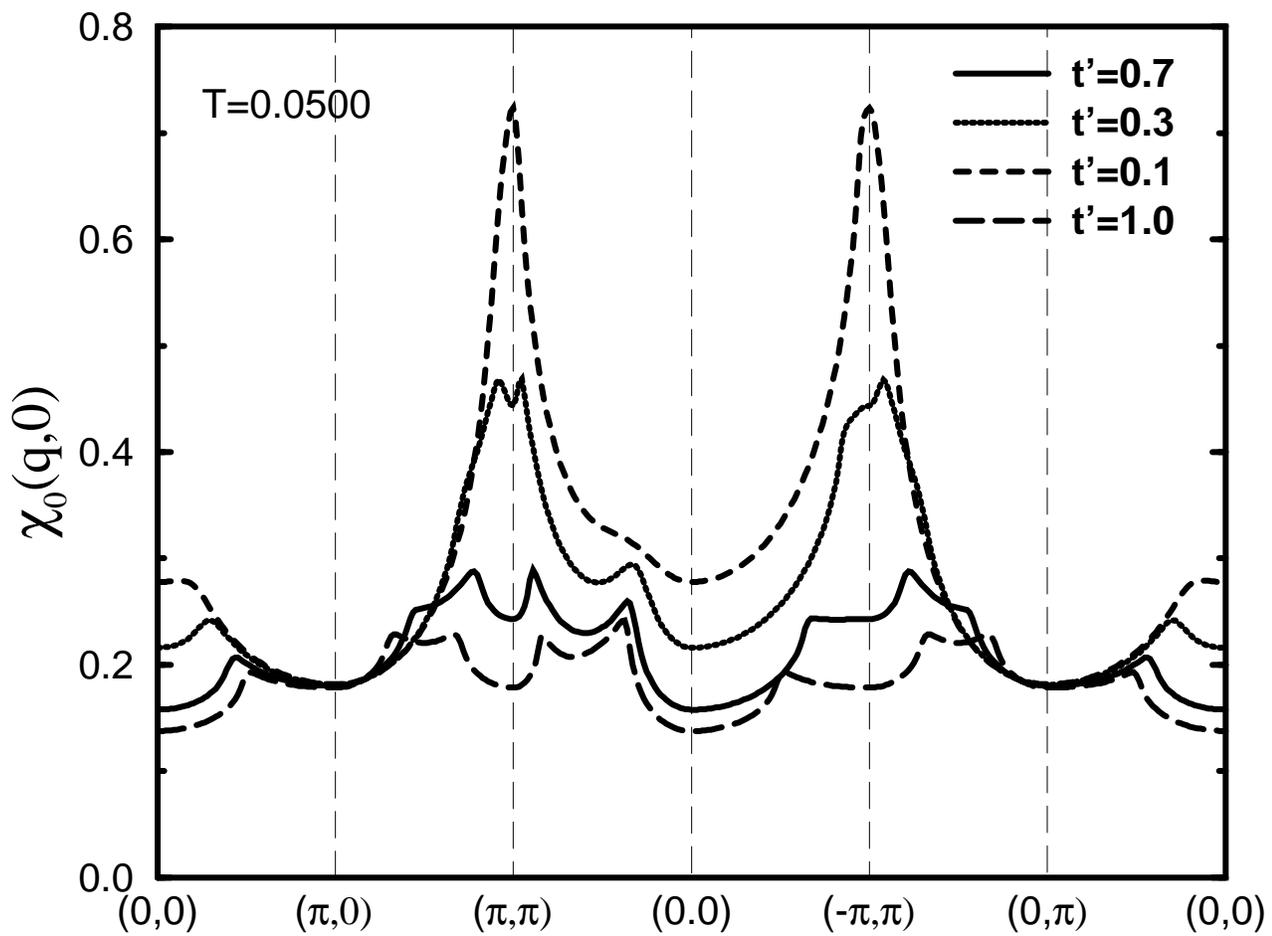
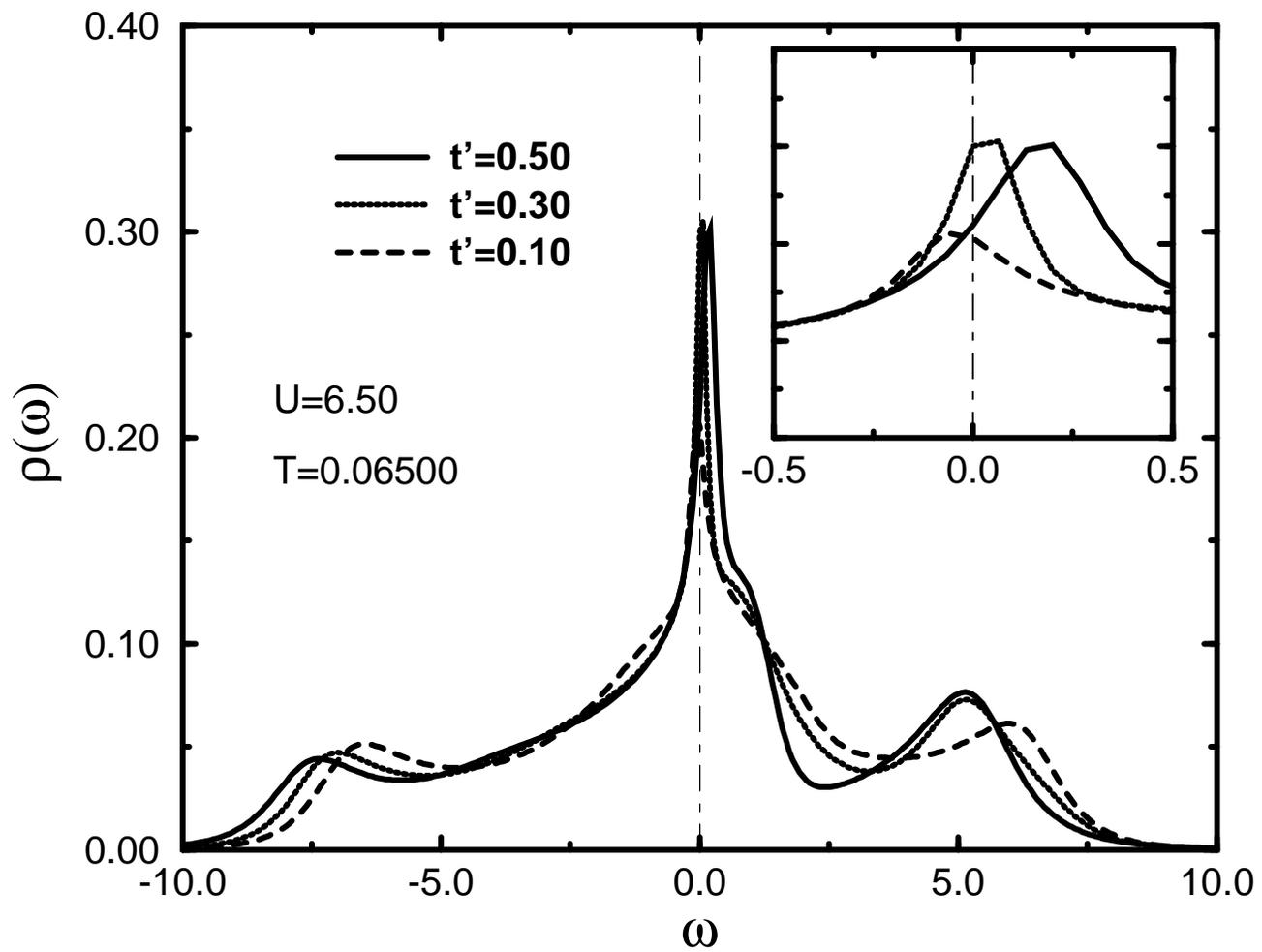
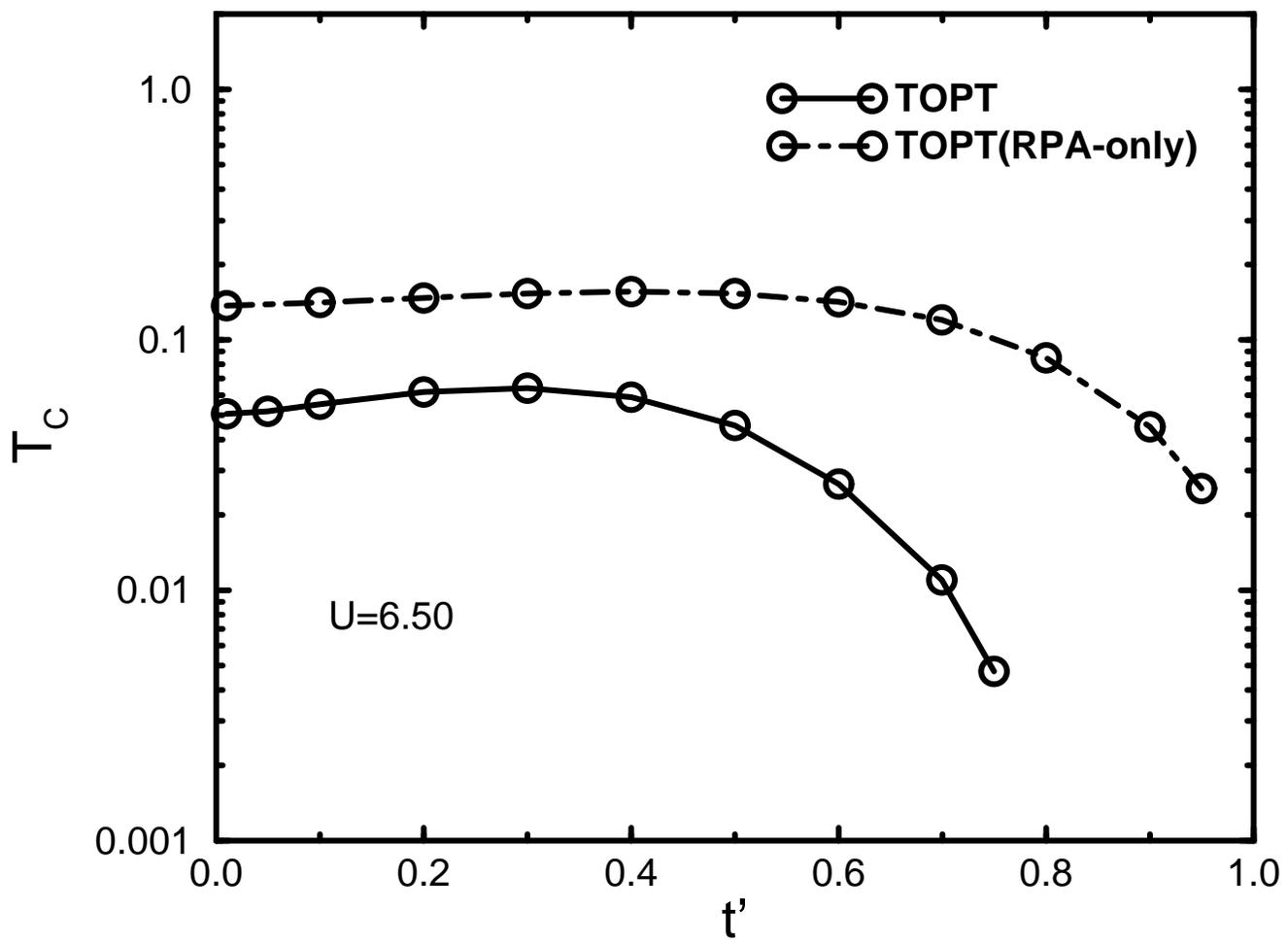
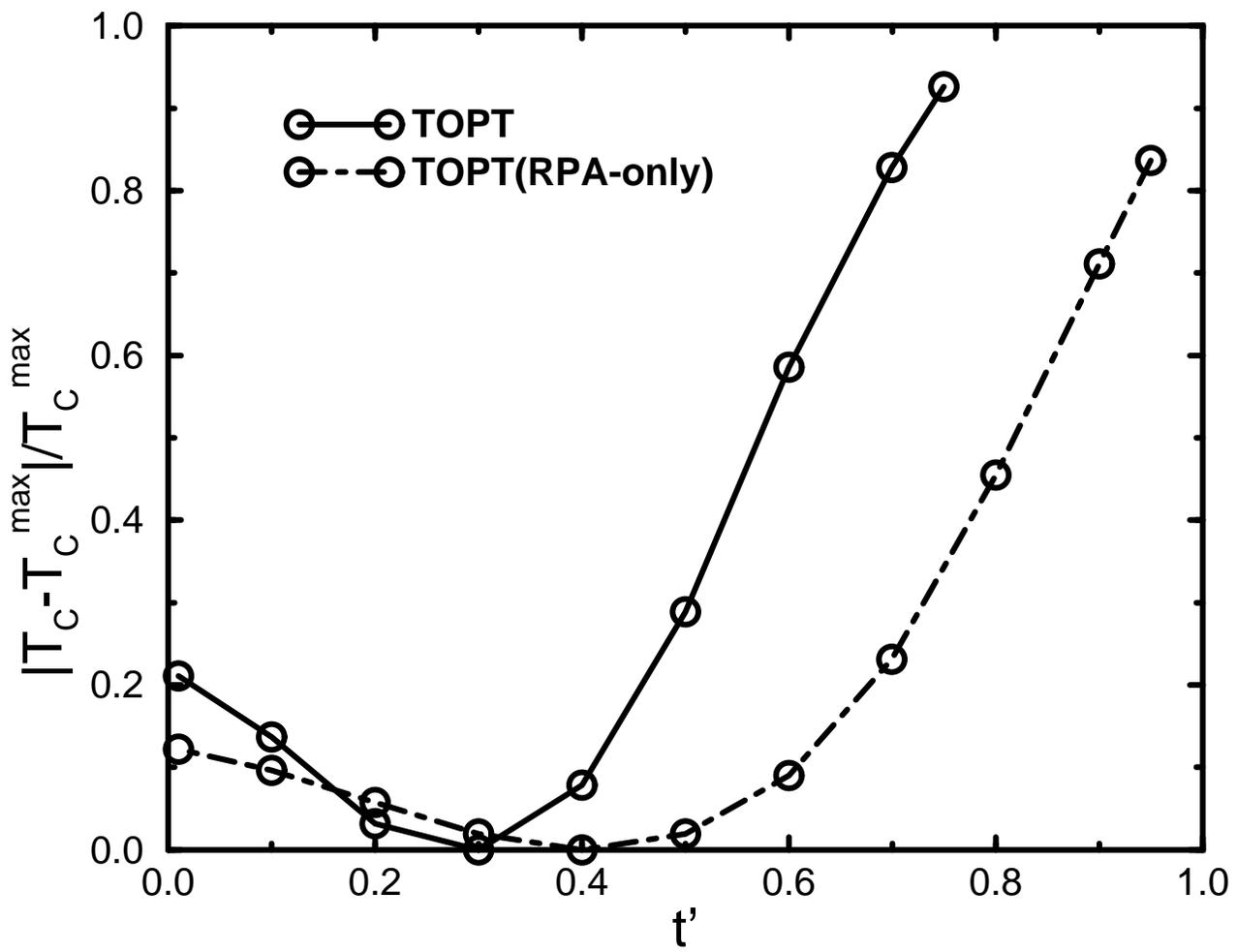
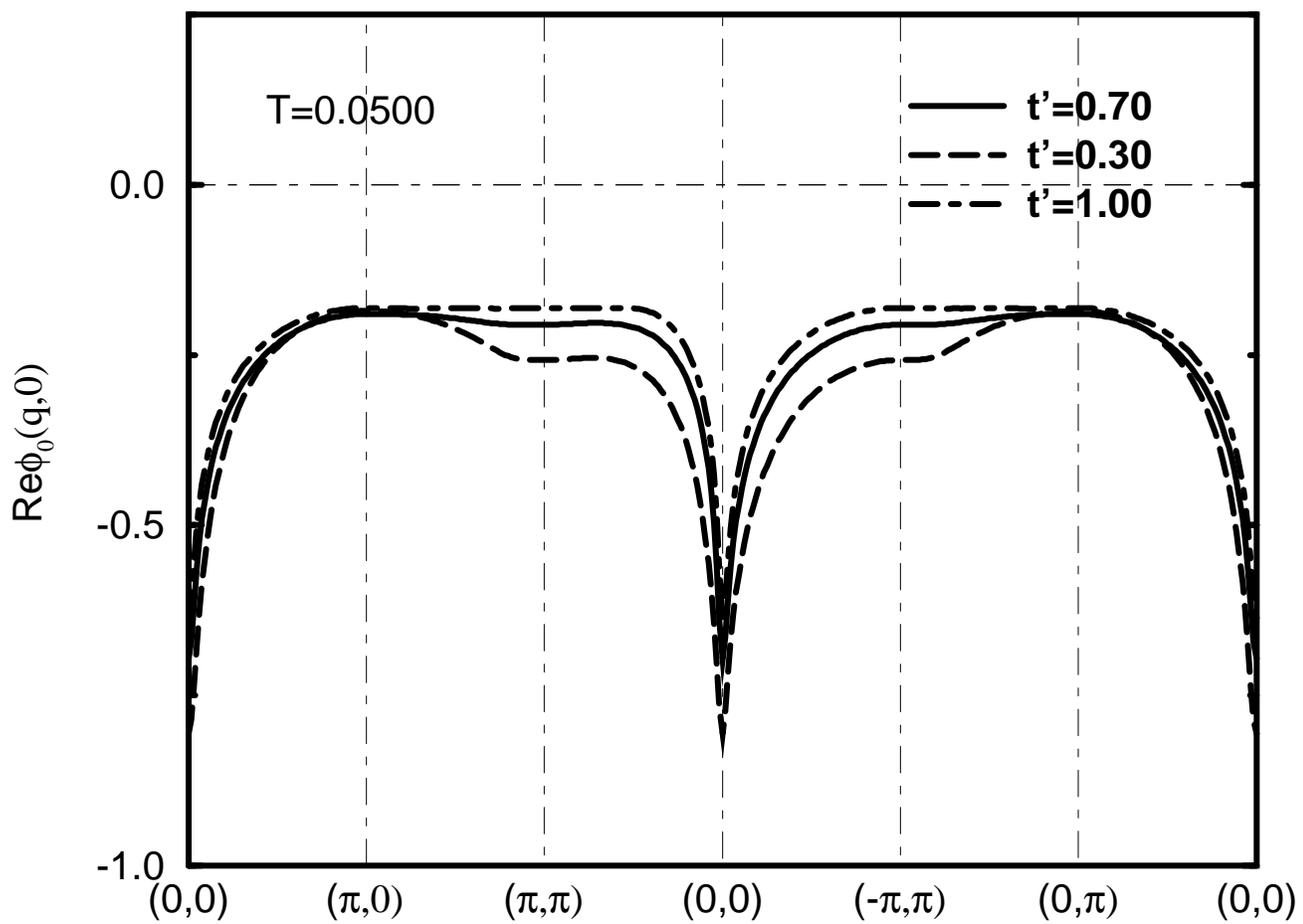
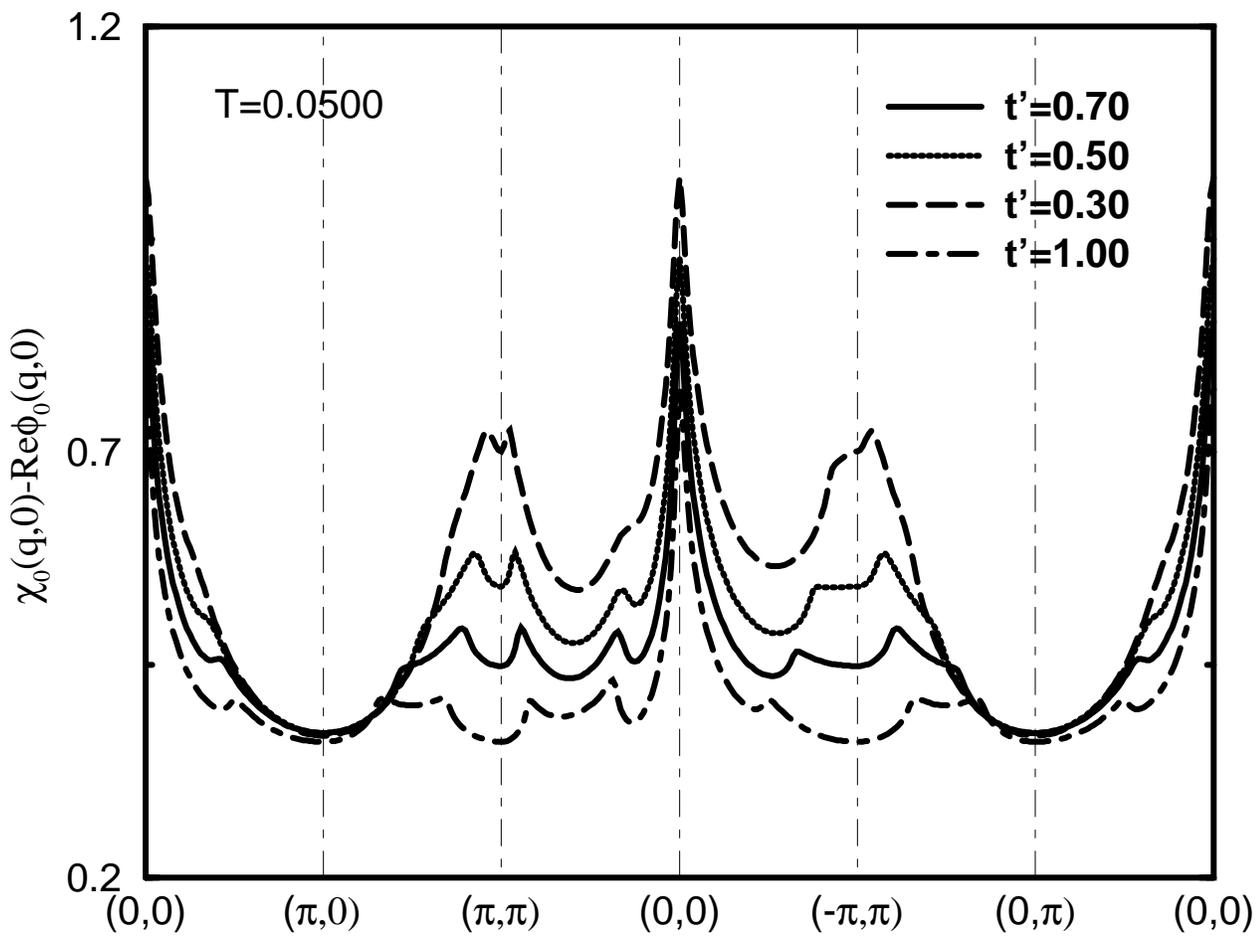
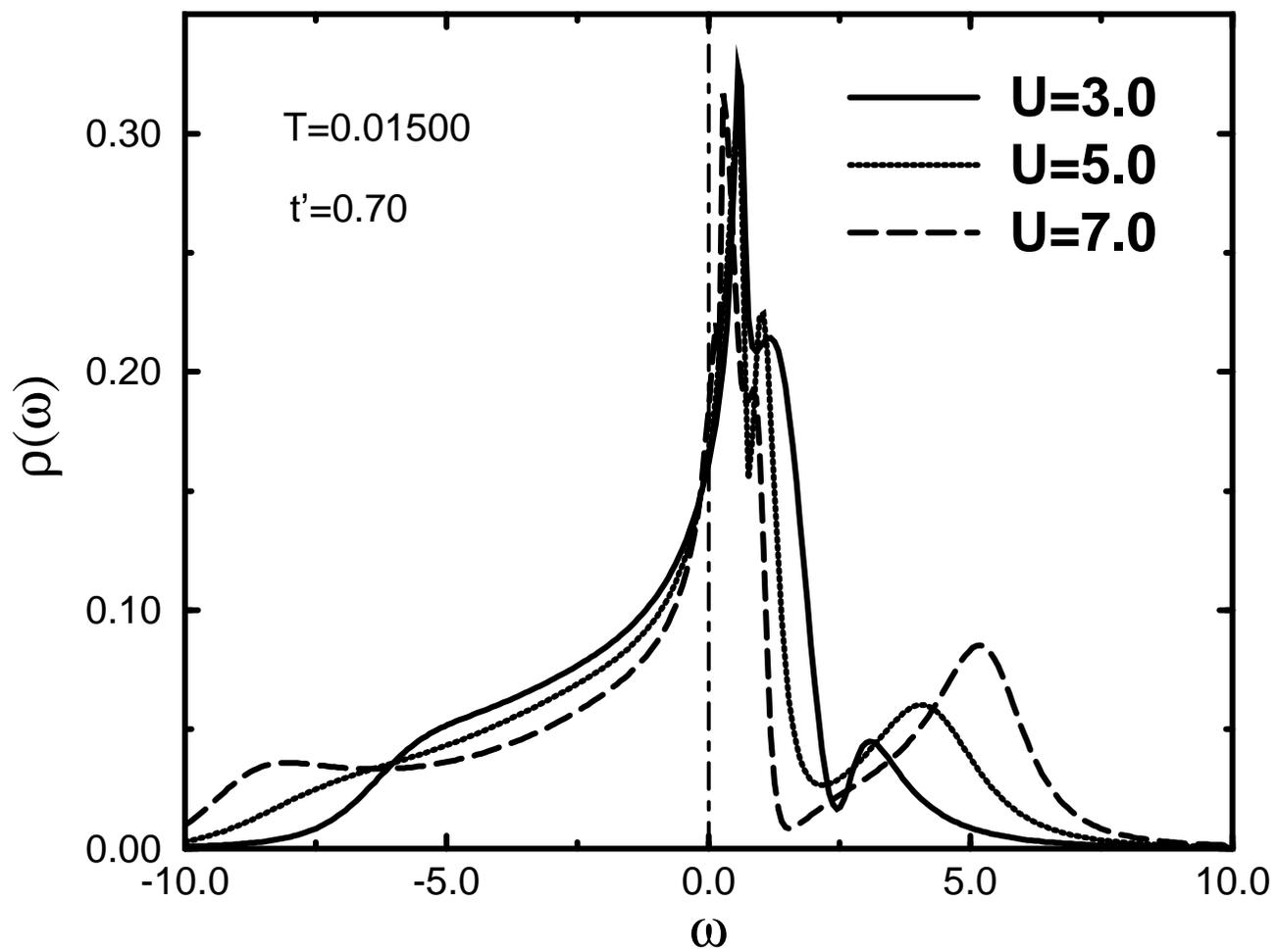
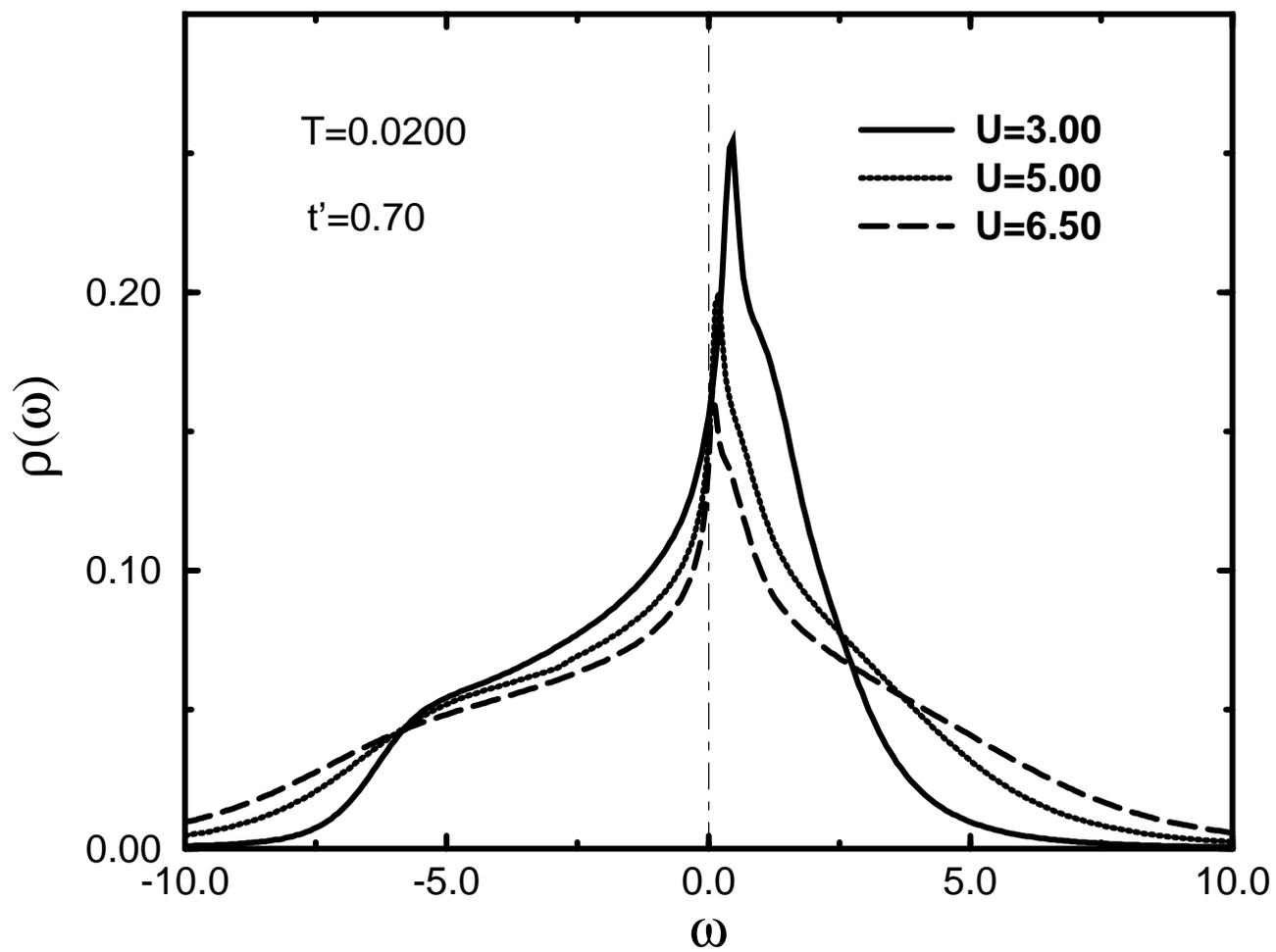
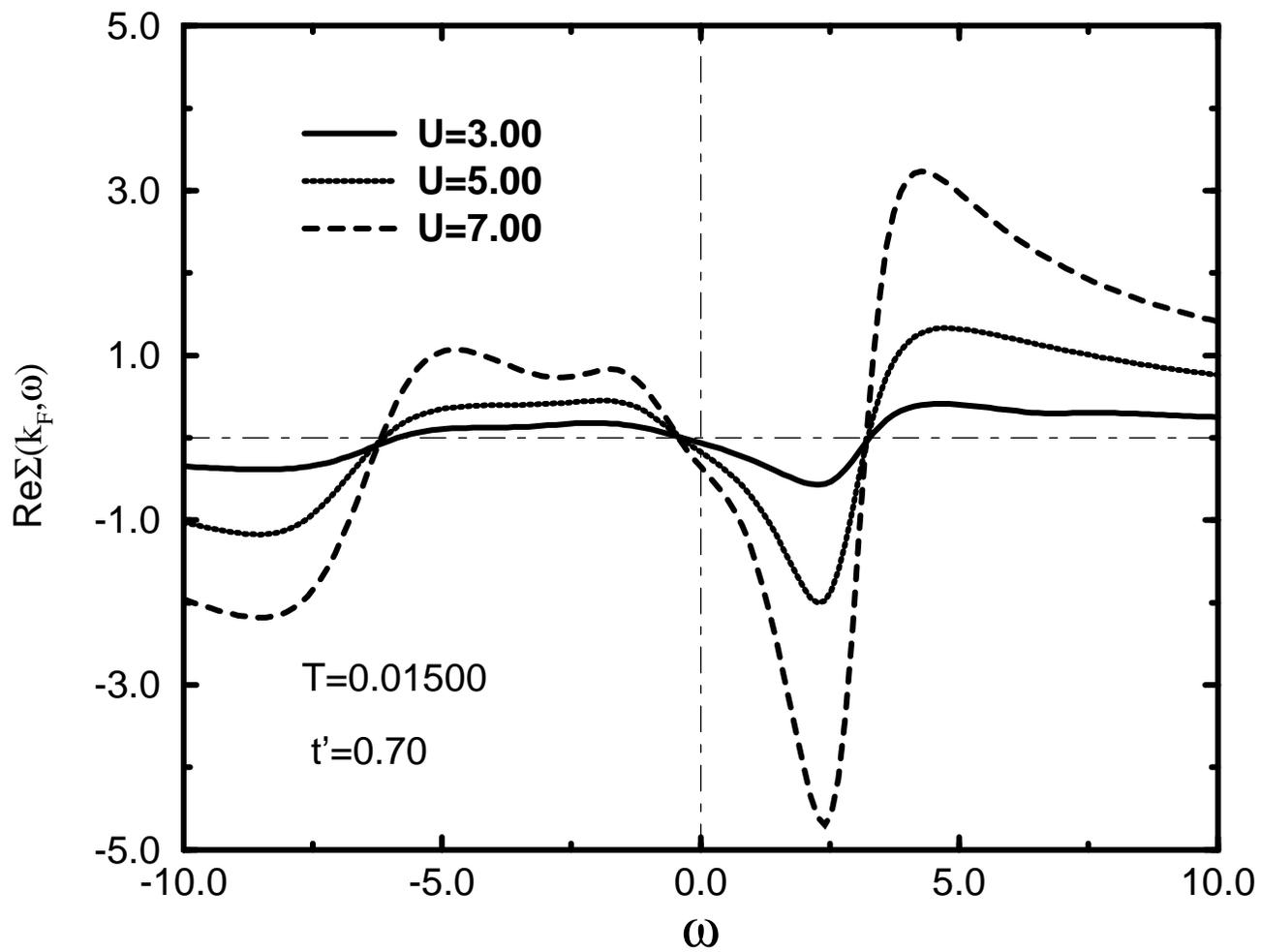
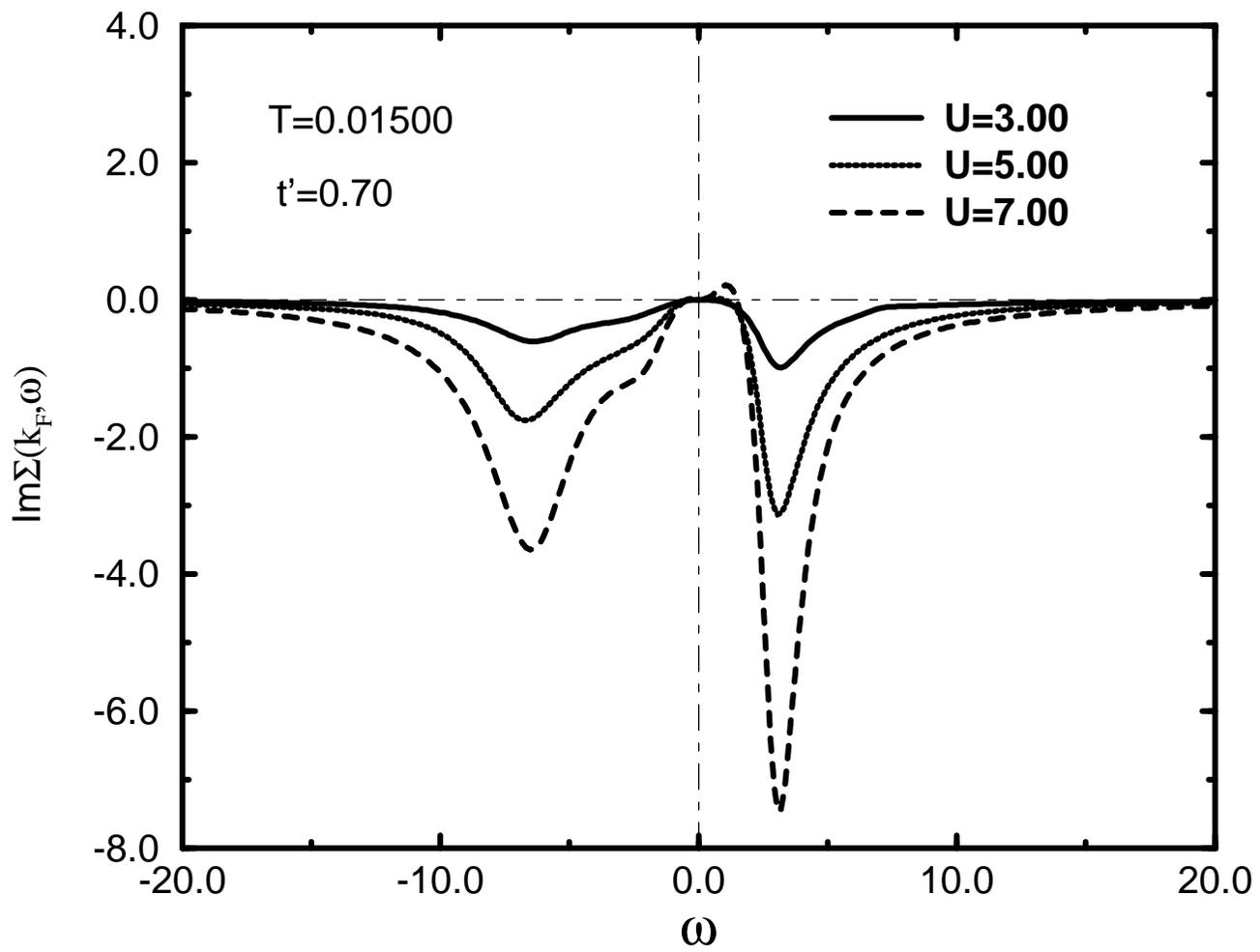
\end{document}